%% file: main.tex
\RequirePackage{fix-cm}
\documentclass{svjour3}                     
\usepackage{graphicx}
\usepackage{amsmath}
\usepackage{amssymb}
\usepackage[mathcal]{eucal}
\input{preamb}

\journalname{General Relativity and Gravitation}
\begin{document}
\title{Analytical time-like geodesics in Schwarzschild space-time}
\titlerunning{Analytical time-like geodesics}        
\author{Uro{\v s} Kosti{\' c}}
\institute{Uro{\v s} Kosti{\' c} \at
              Faculty of Mathematics and Physics, University of Ljubljana\\
			  Jadranska 19, 1000 Ljubljana, Slovenia\\
              \email{uros.kostic@fmf.uni-lj.si}}
\date{Received: 29 August 2011 / Accepted: 15 January 2012}
%
%
\maketitle
\begin{abstract}
Time-like orbits in Schwarzschild space-time are presented and classified in a very transparent and straightforward way into four types. The analytical solutions to orbit, time, and proper time equations are given for all orbit types in the form $\vec{r}=\vec{r}(\lambda)$, $t=t(\chi)$, and $\tau=\tau(\chi)$, where $\lambda$ is the true anomaly and $\chi$ is a parameter along the orbit. A very simple relation between $\lambda$ and $\chi$ is also shown. These solutions are very useful for modelling temporal evolution of transient phenomena near black holes since they are expressed with Jacobi elliptic functions and elliptic integrals, which can be calculated very efficiently and accurately.
\end{abstract}
\section{Introduction\label{sec:introduction}}
\input{intro}
\section{Schwarzschild space-time}
\input{schwarzschild}
\subsection{Types of orbits}
\input{types}
\subsection{Analytical solutions}
\input{solutions}
\subsubsection{Types A and D}
\input{typeA}
\subsubsection{Type B}
\input{typeB}
\subsubsection{Type C}
\input{typeC}
\section{Summary}
\input{conclusion}
%
%
%
\bibliographystyle{spphys}       
\bibliography{biblio}   
\end{document}

%% file: preamb.tex
\newcommand{\beq}{\begin{equation}}
\newcommand{\eq}{\end{equation}}
\newcommand{\bed}{\begin{displaymath}}
\newcommand{\ed}{\end{displaymath}}
\newcommand{\reff}[1]{\mbox{Fig.\hspace{2pt}\ref{#1}}}

\newcommand{\refe}[1]{(\ref{#1})}
\newcommand{\refee}[2]{(\mbox{\ref{#1})\hspace{2pt}--\hspace{2pt}(\ref{#2})}}
\newcommand{\refep}[1]{\mbox{(Eq.\hspace{2pt}\ref{#1})}}

\newcommand{\esp}{\hspace{1cm}}

\newcommand{\lt}{\tilde{l}}

\newcommand{\rp}{r_{\mathrm p}}
\newcommand{\ra}{r_{\mathrm a}}

\newcommand{\ddt}{\widetilde{\mathcal{D}}}
\newcommand{\ddd}{\mathcal{D}}
\newcommand{\dd}{\text{d}}

\newcommand{\cn}{\textrm{cn}}

\newcommand{\am}{\textrm{am}}
\newcommand{\elF}{\textrm{F}}
\newcommand{\elK}{\textrm{K}}
\newcommand{\elE}{\textrm{E}}

\newcommand{\Arth}{\textrm{Artanh}}

\newcommand{\pfi}{p_{\varphi}}
\newcommand{\ptt}{p_t}
\newcommand{\pth}{p_{\theta}}

\newcommand{\pmm}{p_{\mu}}


%% file: intro.tex
When modelling physical phenomena occurring in strong gravitational field of black holes, it is common to work in the Schwarzschild space-time \cite{2002AAS...201.1509R,2003MNRAS.341.1041A,2006ApJ...648..510L,2006ApJ...651.1031S,2006CQGra..23.6503A,2007ApJ...657..415F,2007ApJ...662L..15F,2007ApJ...671.1696S,2008ApJ...679L..93F,2010A&A...521A..67M} . The same applies also for determination of orbital parameters of objects orbiting so close to a black hole that the orbits are affected by relativistic effects, e.g. highly eccentric S stars at the Galactic Centre \cite{2009ApJ...692.1075G,2011MNRAS.411..453I}. For this purpose, an efficient and accurate way for calculating time-like orbits in the Schwarzschild space-time is required. Moreover, if we are interested in time-dependence of these phenomena, a method for solving the time equation is also required to calculate the temporal evolution and the dynamics.

Since in such models, the number of calculations can rapidly increase either because of increasing the number of points, or extending the time of the simulation, or reducing the time-step size, it is very desirable to have a very efficient and accurate method for solving these equations. For example, in a model which includes time-dependant gravitational lensing, it is easy to miss the moment of the strongest lensing when the Einstein ring appears, if the time-step is too large. Consequently, the calculated signal, as received by a distant observer, lacks this distinctive characteristic.

Although the numerical integration or post-Newtonian approximation yield useful results in specific cases, analytical solutions of the orbit and time equation are simpler, more efficient regardless of the accuracy required (as shown by Delva \cite{2010gfps.confE..19D}), and can be used in all cases (weak field limit, strong field limit). The well known work of Chandrasekhar \cite{chandra} and Rauch \cite{rauch}, where the solutions to geodesic equations are expressed in terms of elliptic integrals, has been followed by \v Cade\v z \cite{cadez2,2005PhRvD..72j4024C} and Gomboc \cite{andreja} who inverted the expressions of Chandrasekhar \cite{chandra} and Rauch \cite{rauch} into Jacobi elliptic functions, which no longer contain the branch ambiguity. For light-like geodesics, \v Cade\v z and Kosti\' c \cite{2005PhRvD..72j4024C} presented a very simple and straightforward way of characterizing the orbits which depend only on one parameter, as well as giving analytical solutions to the time equation and a method of determining a light-like geodesic between two arbitrary points (and thus facilitating ray-tracing used in numerical modelling of dynamical phenomena near black holes).

Cruz et al. \cite{2005CQGra..22.1167C} have classified the light-like and time-like geodesics according to the effective potential and found similar analytical solutions as \cite{2005PhRvD..72j4024C}, however they give solutions to time equation only for radial and circular orbits. Hioe and Kuebel \cite{2010PhRvD..81h4017H} present analytical solutions to orbit equations, classify them according to two parameters, and show extensive tables of different values of these parameters for corresponding orbits. They, however, do not give any solution to time equation.

To complement previous work on light-like orbits \cite{2005PhRvD..72j4024C}, the complete analytical solutions of the time-like geodesics and the time equation for all orbit types are presented in this paper: in the form $\vec{r}=\vec{r}(\lambda)$ (where $\lambda$ is the true anomaly) for the radial coordinate $r$, and in the form $t=t(\chi)$ and $\tau=\tau(\chi)$ for time $t$ and proper time $\tau$, respectively, with a very simple relation between $\lambda$ and $\chi$.

%% file: schwarzschild.tex
In Schwarzshild space-time we use Schwarzshild coordinates $t$, $r$, $\theta$, $\varphi$. The Hamiltonian, from which geodesic equations are derived is
\beq
 H=\frac{1}{2}
   \Biggl[
     -\frac{1}{1-\frac{2M}{r}}\ \ptt^2 + \Bigl(1-\frac{2M}{r}\Bigr )p_r^2 +
      \frac{1}{r^2} \Bigl(\pth^2 + \frac{1}{\sin^2\theta} \ \pfi^2\Bigr )
   \Biggr ]\ , \label{hamiltonian}
\eq
where $\pmm$ are canonical momenta and natural units $c=G=1$ are used. The constants of motion are: value of Hamiltonian ($H$) and Lagrangian ($L$), energy $E=p_t$, three components of angular momentum ($\vec{l}$), longitude of periapsis ($\omega$), and time of periapsis passage ($t_p$). For time-like geodesics, the value of Hamiltonian is $H=-1/2$.

\begin{figure}
\sidecaption
\includegraphics[scale=0.16]{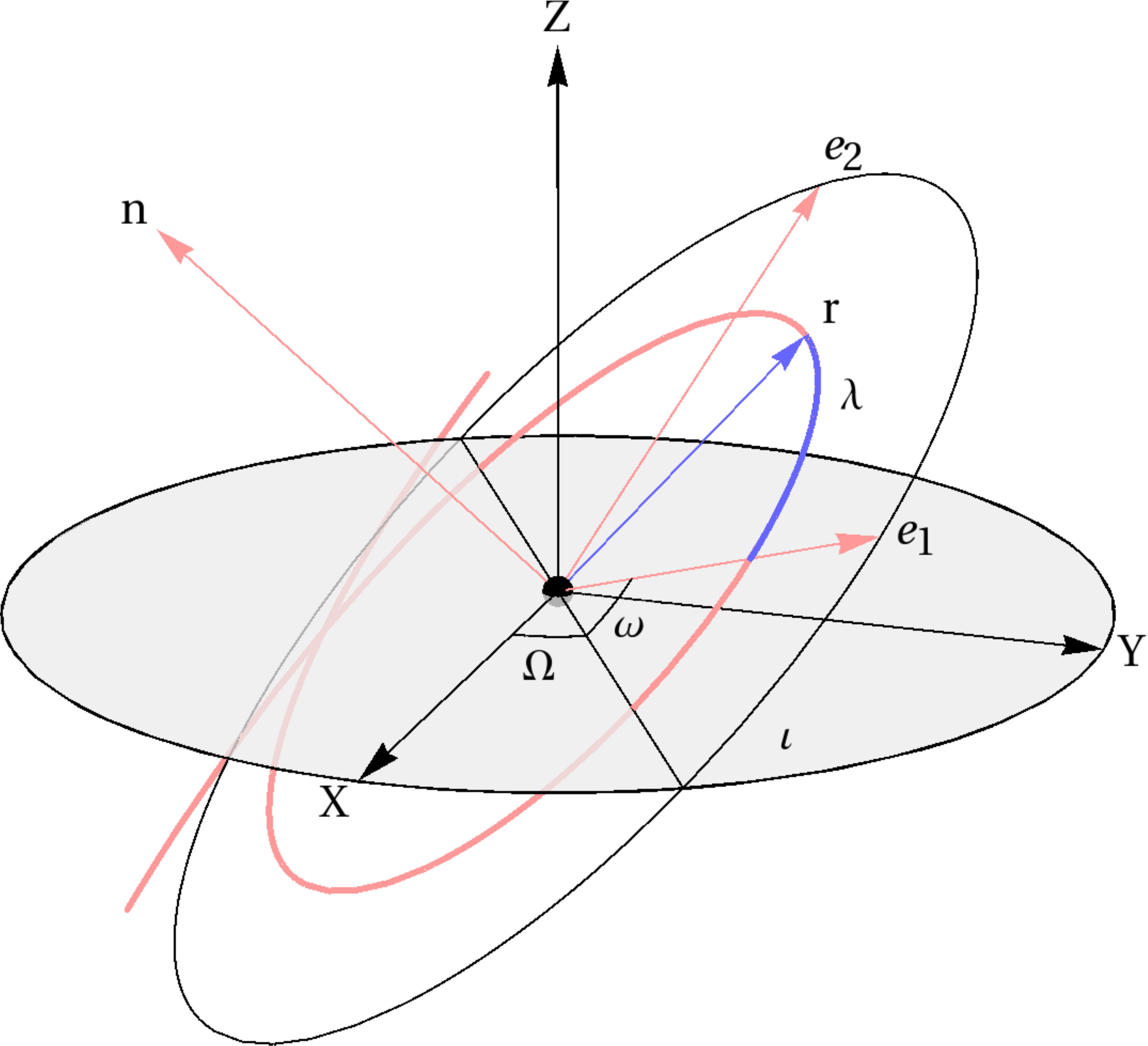}
\caption{The orbital plane in equatorial coordinates: $\hat{n}$ unit normal, $\iota$ inclination, $\Omega$ longitude of the ascending node, $\omega$ longitude of periapsis and $\lambda$ true anomaly.}
\label{fig:l}
\end{figure}

In order to describe the position along the orbit, as well as the orientation of the orbit, we introduce another local inertial (right-handed) orthonormal tetrad $\hat n$, ${\hat e}_1$ and ${\hat e}_2$ as shown in \reff{fig:l}. The vector $\hat n$ is a constant unit vector pointing in the direction of angular momentum ($\vec{l}=l \hat{n}$). The two unit vectors ${\hat e}_1$ and ${\hat e}_2$ in the orbital plane are oriented so that ${\hat e}_1$ points in the direction of initial periapsis, apoapsis or toward the infinity (The choice depends on the orbit type and will be explained further in the text.). The components of these vectors with respect to the local Cartesian coordinate basis are expressed as in \cite{2011AdSpR..47..370D}:
\begin{subequations}
\begin{align}
{\hat e}_1 &=
	\begin{pmatrix}
		\cos\omega \cos\Omega-\cos\iota\sin \omega\sin\Omega\\
    	\cos\omega\sin\Omega+\cos \iota \sin \omega\cos\Omega\\
		\sin\iota\sin\omega
	\end{pmatrix}
\\
{\hat e}_2 &=
	\begin{pmatrix}
		-\sin\omega \cos\Omega-\cos\iota\cos \omega\sin\Omega\\
		-\sin\omega\sin\Omega+\cos \iota \cos \omega\cos\Omega\\
		\sin\iota\cos\omega
	\end{pmatrix}
\\
\hat n &= \left (\sin\iota\sin\Omega,-\sin\iota\cos\Omega, \cos\iota\right )
\label{eq:vrt_kol_komponente}
\end{align}
\end{subequations}
where $\Omega$ is the longitude of the ascending node and $\iota$ is the inclination of the orbit with respect to the $X-Y$ plane (see \reff{fig:l}).

By introducing a dimensionless variable
\beq
	u=\frac{2M}{r}
\eq
and two dimensionless constants of motion related to orbital energy and orbital angular momentum \cite{andreja}:
\beq
	a = \frac{2ME}{l} \esp \mathrm{and} \esp b = 2H\left(\frac{2M}{l}\right)^2
\eq
one can derive the differential orbit equation:
\beq
  \frac{\dd u}{\dd\lambda} = \pm\sqrt{a^2-u^2(1-u)+b(1-u)} \ ,
\label{eq:orbit_diff}
\eq
where $\lambda$ is the true anomaly. As functions of $u$, time and proper time obey the following differential equations:
\begin{subequations}
\begin{align}
	\frac{\dd t}{\dd u} & = \frac{2Ma}{	u^2 (1-u)\sqrt{a^2 - u^2(1-u) + b(1-u)}	}\label{eq:times1}\\
	\frac{\dd \tau}{\dd u} & = \frac{2Ma}{E}\frac{1}{	u^2\sqrt{a^2 - u^2(1-u) + b(1-u)}	}  .
\label{eq:times}
\end{align}
\end{subequations}
After \refe{eq:orbit_diff} is solved for $u$ as a function of $\lambda$, the orbit equation is written in vector form as
\beq
\vec{r}(\lambda) = \frac{2M}{u(\lambda)}(\hat{e}_1 \cos \lambda + \hat{e}_2 \sin \lambda)\ .
\label{eq:orbit_vec}
\eq

Solutions depend on the type of orbit, e.g. closed, scattering or plunging, and in the following section we present them for all types of time-like geodesics.\footnote{The differential equations \refe{eq:orbit_diff} and \refe{eq:times1} are formally the same for light-like geodesics \cite{2005PhRvD..72j4024C} (for light-like geodesics take $b=0$).}

%% file: types.tex
\begin{figure}
\includegraphics[width=8.5cm]{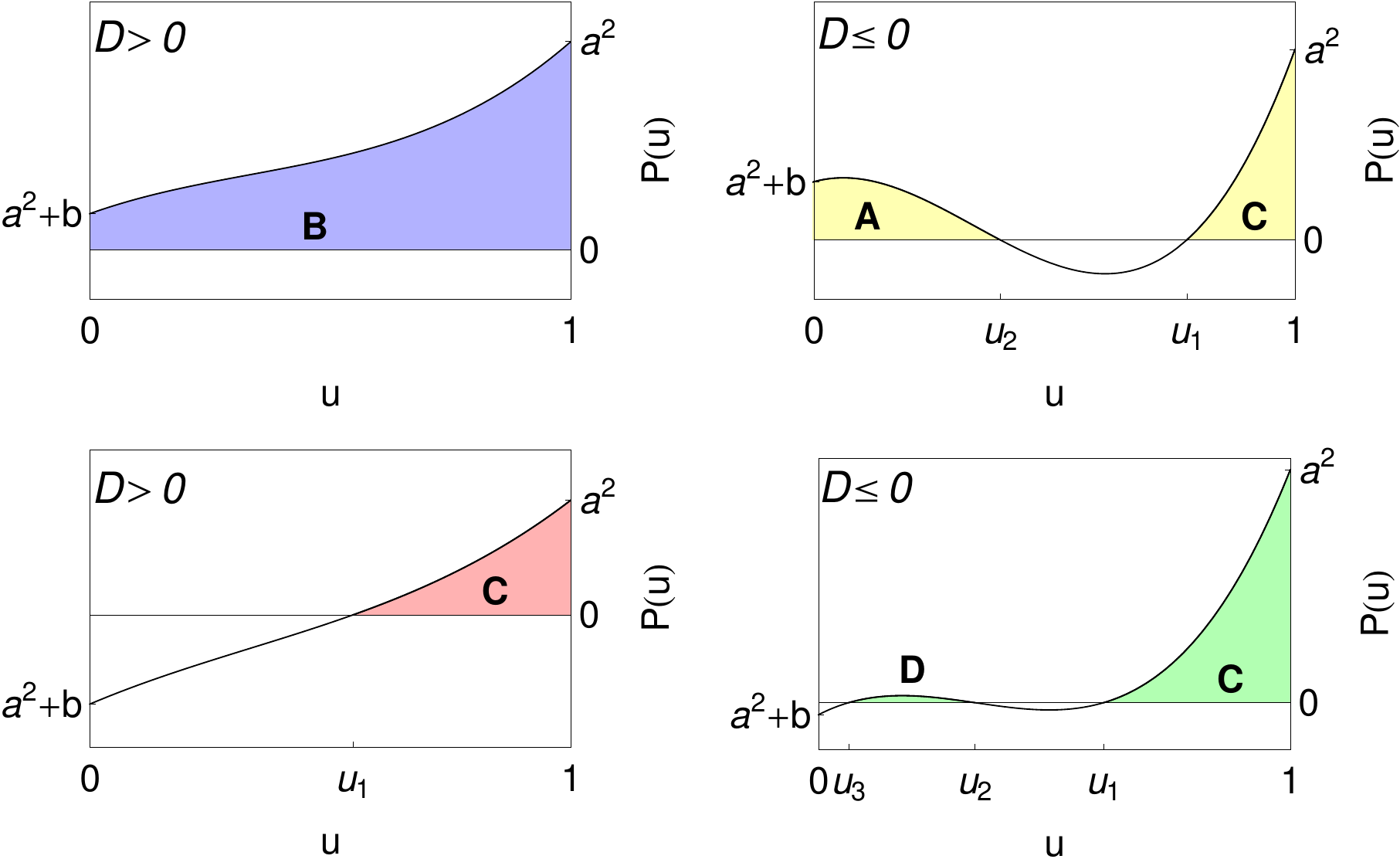}
\caption{The polynomial $P(u)$ and distribution of its roots in the interval $u \in [0,1]$. Orbits exist only where $P(u)\geq 0$, shown in colours. Corresponding orbit types are marked with letters \textbf{\emph{A}}, \textbf{\emph{B}}, \textbf{\emph{C}} or \textbf{\emph{D}} and the roots are marked with $u_1$, $u_2$ and $u_3$. The sign of the discriminant $D$ is also noted.}
\label{fig:tipi}
\end{figure}

\label{sec:types}
Marking the polynomial in \refe{eq:orbit_diff} with $P(u)=a^2-u^2(1-u)+b(1-u)$, the solutions to \refee{eq:orbit_diff}{eq:times} exist only on intervals where $P(u)\geq 0$. This polynomial has three roots, while the discriminant $D$, which is defined as:
\begin{align}
	\alpha & = 1 - 9b - \frac{27}{2}a^2\\
	 \beta & = -1 -3b\\
			 D & = \alpha^2 + \beta^3\ ,
\end{align}
determines the nature of these roots (i.e. the number of real/complex roots). Since orbits extend at most from $u=0$ to $u=1$, only roots on this interval are of interest. In \reff{fig:tipi}, the polynomial $P(u)$ is plotted for all the four possible orbit types (according to the number of roots in the interval $u\in (0,1)$).

The classification of orbits is more intuitive when it is done with respect to the effective potential $V$ defined as \cite{1973grav.book.....M}
\beq
V = \sqrt{(1-u)(1 + \lt^2 u^2)}\ ,
\label{eq:potencial}
\eq
where $\lt=l/2M$ is the reduced angular momentum. Unlike in Keplerian case, the effective potential gains a maximum $V_{max}$
\beq
V_{max} = \left (
				\left(	1	+	\frac{\lt^2}{(\lt^2 -\lt\sqrt{\lt^2 - 3})^2}
				\right)
				\left( 1	-	\frac{1}{\lt^2 -\lt\sqrt{\lt^2 - 3}}
				\right)
			\right )^{1/2}
\label{eq:Vmax}
\eq
at radius $r_{max}$\footnote{Note that $r_{max}$ is \emph{not} the maximal radius an orbit can extend to, but the radius where $V_{max} = V(r_{max})$.}
\beq
r_{max} = 2M\lt\left(\lt - \sqrt{\lt^2 - 3}\right)\ .
\label{eq:rmax}
\eq
Clearly, the maximum exists only for $\lt>\sqrt{3}$. The existence of $V_{max}$ greatly affects the nature of orbits,\footnote{Obviously, the orbits can exist only for $E\geq V$.} especially if the orbital energy $E$ is $E\sim V_{max}$: such orbits can wind around the black hole at $r_{max}$ several times before continuing either away from or towards the black hole, and do not exist in case of Newtonian potential. If $\lt=\sqrt{3}$, the maximum (and the minimum) of the potential disappears at $r_{max} = 6M$, which is the radius of the last stable circular orbit.

\begin{figure}
\sidecaption
\includegraphics[width=6.5cm]{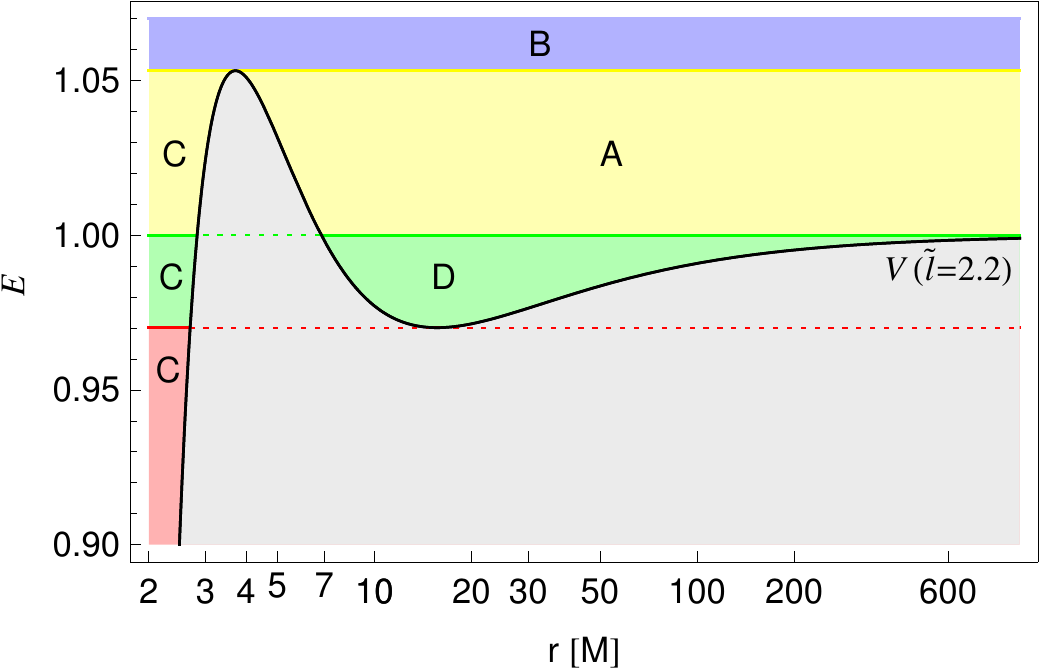}
\caption{The effective potential $V$ for time-like geodesics \refep{eq:potencial} for $\lt=2.2$. By choosing appropriate value of $E$, orbits of any type can be constructed: orbits with $E>V_{max}$ (blue area) are of type B (plunging orbits), orbits with $1\leq E \leq V_{max}$ (yellow area) are either of type A (scattering orbits) or of type C (near orbits), orbits with $V_{min} \leq E < 1$ (green area) are either of type D (bound orbits) or of type C, and orbits with $E<V_{min}$ (red area) are only of type C. Note that for $V_{min} \leq E \leq V_{max}$ the discriminant is $D\leq 0$, and $D>0$ otherwise.}
\label{fig:potencial}
\end{figure}

\begin{figure}
\includegraphics[width=\textwidth]{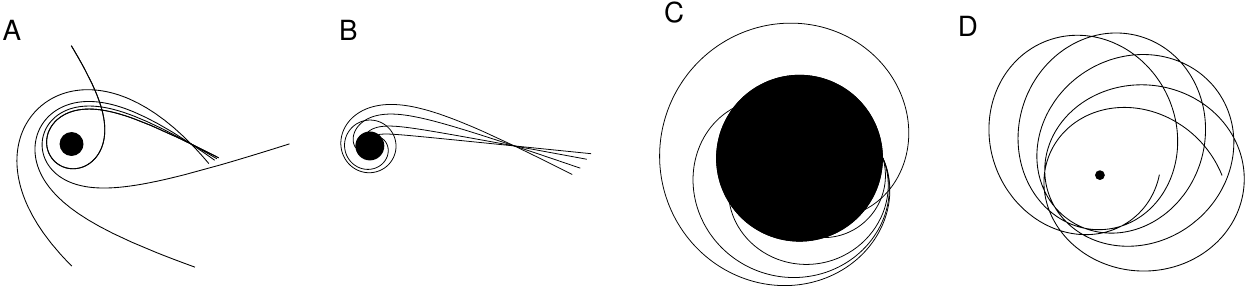}
\caption{Time-like geodesics with $\lt=2.32379$ (for types \emph{A}, \emph{B}, and \emph{C}) and $\lt=3$ (for ype \emph{D}). From left to right: Orbits of type \emph{A} with $E \in \lbrace 1.0001, 1.035, 1.06, 1.083 \rbrace$, orbits of type \emph{B} with $E \in \lbrace 1.0887, 1.2, 1.6, 2.5\rbrace$, orbits of type \emph{C} with $E \in \lbrace 0.7, 0.97372899, 1.05, 1.086\rbrace$, an orbit of type \emph{D} with $E = 0.988$. The radius of the black circle is the Schwarzschild radius.}
\label{fig:orbitsABCD}
\end{figure}

The effective potential and corresponding orbit types are shown in \reff{fig:potencial}.
The four types of orbits for massive particles have the following properties:
\begin{itemize}
\item[-] type \emph{A}: scattering orbits with both endpoints at infinity. Scattering orbits can never extend below $r=3M$.
\item[-] type \emph{B}: plunging orbits with one end at infinity and the other behind the horizon,
\item[-] type \emph{C}: near orbits with both ends behind the horizon of the black hole.
\item[-] type \emph{D}: bound orbits. Highly eccentric orbits can never reach below $r=4M$ while circular orbits can never reach below $r=6M$.\footnote{Highly eccentric orbits are orbits with energy $E\lesssim 1$ (which makes the orbits almost parabolic). From equations \refe{eq:Vmax} and \refe{eq:rmax} it follows that for type D orbits, $r_{max}$ is the smallest if $V_{max} \approx E \approx 1$, which happens for $\lt=2$ at $r_{max}\approx 4 M$. In this case, $r_{max}$ corresponds to the periapsis distance.}
\end{itemize}
Some typical examples of all types are shown in \reff{fig:orbitsABCD}. Note that, only if $E\approx V_{max}$, then $r_{max}$ corresponds to the radius of periapsis for type A and D orbits, and apoapsis for type C orbits.

\begin{figure}
\sidecaption
\includegraphics[width=6.5cm]{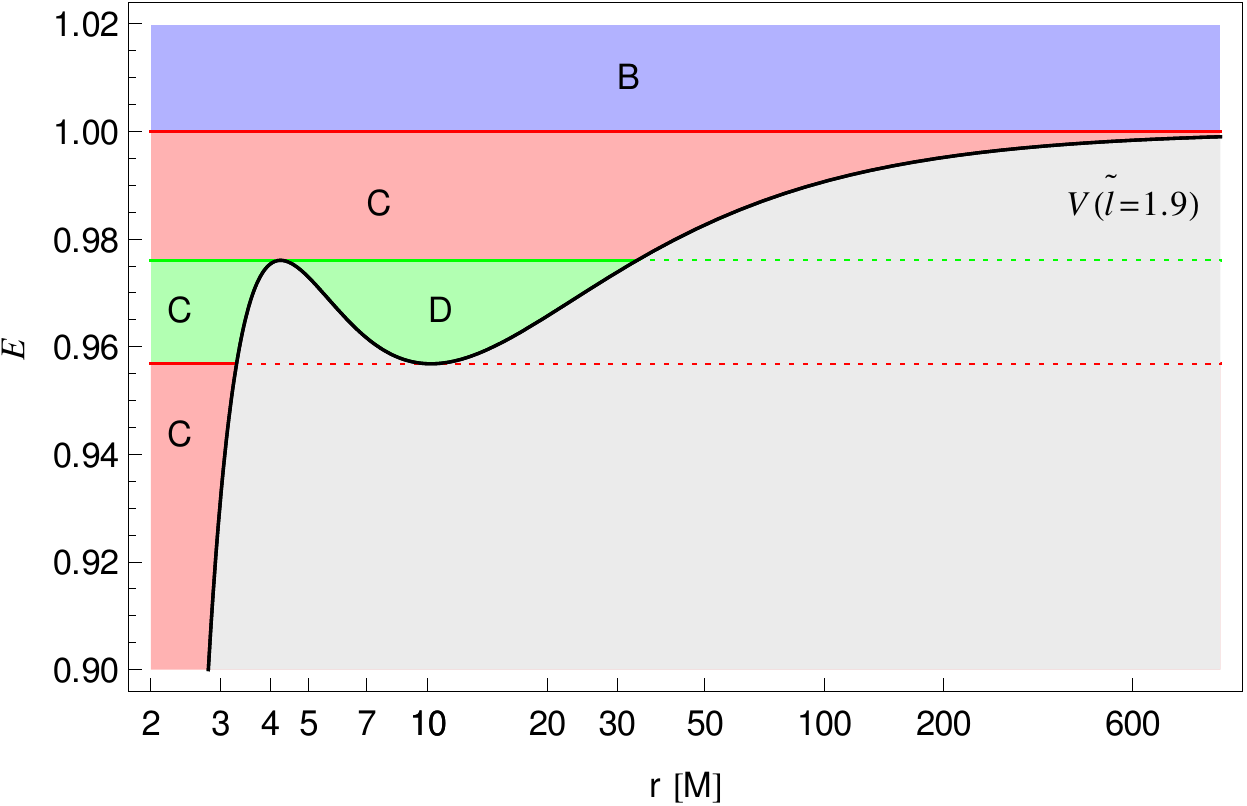}
\caption{The effective potential $V$ for time-like geodesics \refep{eq:potencial} for $\lt=1.9$. By choosing appropriate value of $E$, only orbits of type B, C, and D can be constructed: orbits with $E>1$ (blue area) are of type B (plunging orbits), orbits with $ V_{max}\leq E \leq 1$ (top red area) are of type C (near orbits), orbits with $V_{min} \leq E < V_{max}$ (green area) are either of type D (bound orbits) or of type C, and orbits with $E<V_{min}$ (bottom red area) are of type C. Note that for $V_{min} \leq E \leq V_{max}$ the discriminant is $D\leq 0$, and $D>0$ otherwise.}
\label{fig:potencial2}
\end{figure}

\begin{figure}
\sidecaption
\includegraphics[width=4cm]{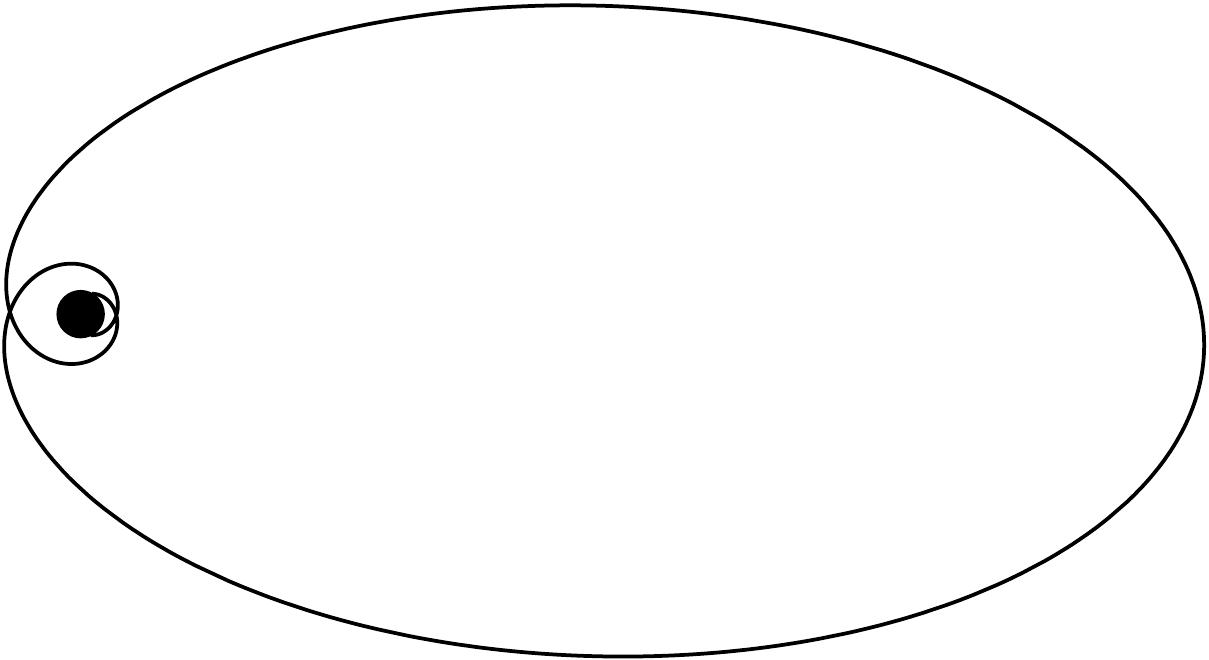}
\caption{Orbit of type C for $\lt=1.9$ and $E=0.99$. The radius of the black circle is the Schwarzschild radius.}
\label{fig:orbitC_big}
\end{figure}

If, however, $\lt<2$ then $V_{max}<1$ (see \refe{eq:Vmax}) and consequently, orbits of type A no longer exist, as shown in \reff{fig:potencial2}. Moreover, while orbits of type C still have both endpoints behind the horizon of the black hole, they can extend to infinity for $E\rightarrow 1$. An example of such extended type C orbit for $\lt=1.9$ is in \reff{fig:orbitC_big}. Furthermore, if $\lt$ is lowered below $\sqrt{3}$, also type D orbits no longer exist, and only orbits of type B and C remain.

Radial and circular orbits can be considered as special cases of type B and D orbits, respectively. The corresponding equations and parameters for radial orbits are: $\dot{t}= E/(1-u)$, $\dot{r} = \sqrt{-(1-u)+E^2}$ with zero angular momentum $\lt=0$, while for circular orbits they are: $\dot{r} = 0$, $\dot{\lambda} = l/r^2 = const.$, $\dot{t} = E/(1-u) = const.$, with energy $E=V_{min}(\lt)$ where $V_{min}(\lt)$ is the minimum of the effective potential \refe{eq:potencial}.

%% file: solutions.tex
\label{sec:solutions}
The solutions of the equations \refee{eq:orbit_diff}{eq:times} are the following.

%% file: typeA.tex
In this case, the polynomial $P(u)$ has either two ($u_1$ and $u_2$) or three ($u_1$, $u_2$, and $u_3$) real roots on the interval $(0,1)$ which can be elegantly expressed with the constants $\alpha$, $\beta$, and $D$ using Cardan's formula \cite{nickalls} by introducing two more intermediary constants $\ddd$ and $\psi$ \cite{andreja}:
\begin{align}
	|\ddd| & = \sqrt{-\beta}\label{dd}\\
	\psi & = 2 \arctan \left( \frac{\sqrt{-D}}{\alpha + \sqrt{-\beta^3}} \right)\ .\label{eq:psi}
\end{align}
In terms of these, the roots can be written in the trigonometric form:
\begin{subequations}
\label{sub:par_a}
\begin{align}
    u_1 & = \frac{1}{3}\Bigl( 1+2|\ddd|\cos\frac{\psi}{3}  \Bigr) \label{u1m}\\
    u_2 & = \frac{1}{3}\Bigl( 1+2|\ddd|\cos\frac{\psi-2\pi}{3}  \Bigr)\label{u2m}\\
    u_3 & = \frac{1}{3}\Bigl( 1+2|\ddd|\cos\frac{\psi+2\pi}{3}  \Bigr) \label{u3m}\ .
\end{align}
\end{subequations}
These roots can be associated to the radius of periapsis $\rp=2M/u_2$ (types \emph{A}, \emph{D}) and apoapsis $r_a=2M/u_3$ (type \emph{D} only). Since the argument of $arctan$ in \refe{eq:psi} is positive, it follows that $0\leq \psi\leq \pi$, therefore $u_1 > u_2 > u_3$.

Using the substitution \cite{elipticni}
\beq
	u(\chi) = u_2 - (u_2 - u_3)\cos^2 \chi \label{eq:u_chi}\ ,
\eq
equations \refee{eq:orbit_diff}{eq:times} are transformed into Legendre form of elliptic integrals and integrated to obtain orbital variables $\lambda$, $t$, and $\tau$ as functions of $\chi$:
\begin{align}
	\lambda (\chi) & = n \left( \elF (\chi|m) - \elK (m)	\right)\label{eq:lambda_chi}\\
\begin{split}
	t(\chi) & = \frac{2na}{u_3^2}\Biggl[
	\left(
	1+u_3+\frac{  n_1^2-m  }{  2(m-n_1)(n_1-1)  }
	\right)\Pi(n_1;\chi|m) +
	\frac{u_3^2}{1-u_3}\Pi(n_2;\chi|m) \\
	  &\quad +
	\frac{n_1/2}{(m-n_1)(n_1-1)}
	\left(
	\elE(\chi|m)-\Bigl(1-\frac{m}{n_1}\Bigr)\elF(\chi|m)\right . \\
	&\left . \quad -
	\frac{
	n_1\sin 2\chi\sqrt{1-m\sin^2 \chi}
	}{
	2(1-n_1\sin^2 \chi)
	}
	\right)
	\Biggr]
	\label{resiteva}
\end{split}\\
	\tau (\chi) & = \frac{1}{E} t(\chi)-
					\frac{2n}{\lt u_3 }\left( \Pi (n_1;\chi | m) + 
													\frac{u_3}{1 - u_3} \Pi(n_2;\chi|m)
												\right)\ , \label{eq:tau}
\end{align}
where:
\begin{subequations}
\begin{align}
 	m & = \frac{u_2-u_3}{u_1-u_3}\\
	n & = \frac{2}{\sqrt{u_1-u_3}}\\
	n_1 & = 1 - \frac{u_2}{u_3}\\
	n_2 & = \frac{u_2 - u_3}{1 - u_3}\ .
\end{align}
\end{subequations}
Inverting \refe{eq:lambda_chi} by $\chi (\lambda) = \am (	\elK (m) + \lambda/n\ | m )$ and using \refe{eq:u_chi} one can also write the solution to the orbit equation \refe{eq:orbit_diff} as a function of true anomaly in the form:
\beq
	u(\lambda) = u_2-(u_2-u_3)\cn^2
		  \bigl(
		  \elK(m) + \frac{\lambda}{n}|m
	\bigr)\ . \label{orbitaADm}
\eq

For type \emph{D} orbits, both $\lambda$ and $\chi$ can go from $-\infty$ to $+\infty$. For type \emph{A}, the values of $\chi$ are in the interval $\chi \in (\chi_{min},\chi_{max})$, where $\chi_{min}=\arccos(\sqrt{u_2/(u_2-u_3)})$ and $\chi_{max}=\arccos(-\sqrt{u_2/(u_2-u_3)})$, while the values of $\lambda$ are in the interval $\lambda/n \in (\elF(\chi_{min}|m) - \elK(m),\elF(\chi_{max}|m) - \elK(m))$. The values of $\chi$ at periapsis and apoapsis are $\pi/2$ and $0$ respectively, while $\lambda = 0$ at periapsis. Definitions of elliptic integrals and functions are from Wolfram \cite{mathematica}.

\begin{figure}
\includegraphics[width=\textwidth]{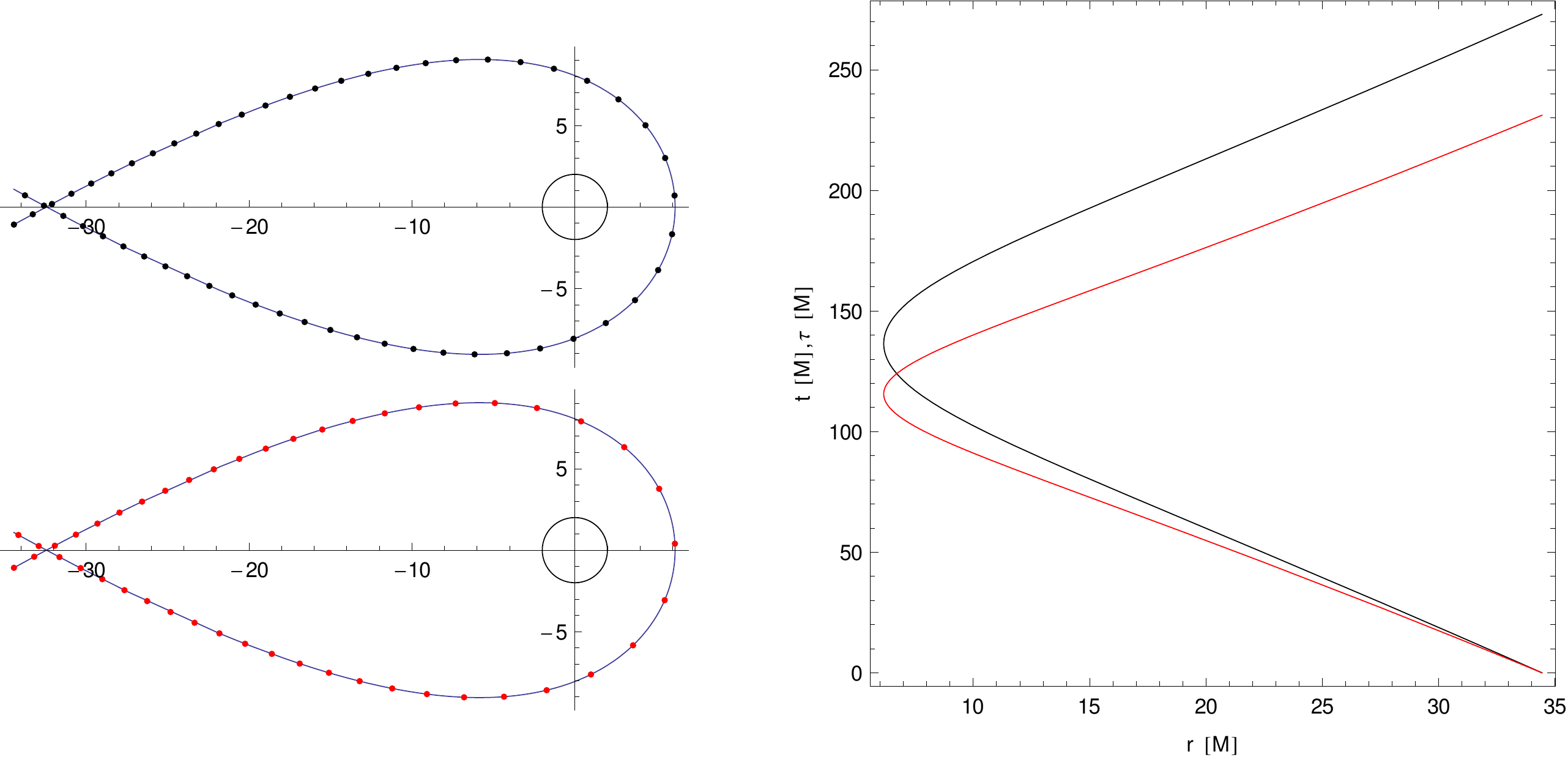}
\caption{Left: Orbit of type A with $E=1.01$ and $\lt=2.2$. Black and red dots correspond to points at time intervals $\Delta t=5 M$ and proper time intervals $\Delta\tau = 5M$, respectively. The black circle represents the Schwarzschild radius. Right: Time (black) and proper time (red) as a function of coordinate $r$, measured from the initial point at $r=34 M$.}
\label{fig:outA}
\end{figure}

\begin{figure}
\includegraphics[width=\textwidth]{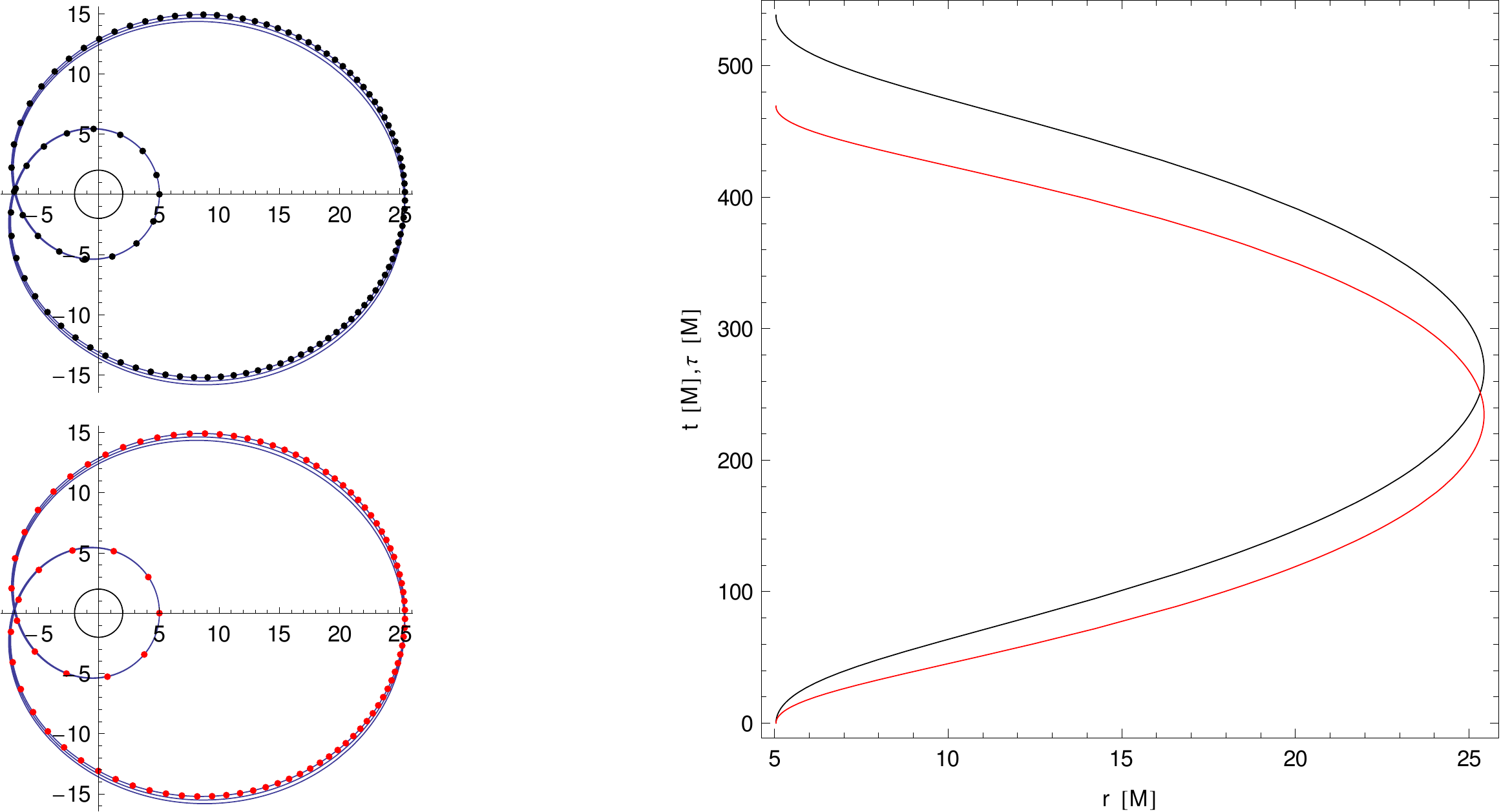}
\caption{Left: Orbit of type D with $E=0.9704$ and $\lt=1.888$. Black and red dots correspond to points at time intervals $\Delta t=5 M$ and proper time intervals $\Delta\tau = 5M$, respectively. The black circle represents the Schwarzschild radius. Right: Time (black) and proper time (red) as a function of coordinate $r$, measured from the initial point at $r=\rp\sim 5.045 M$.}
\label{fig:outD}
\end{figure}

In Figures \ref{fig:outA} and \ref{fig:outD}, an example of solutions for a type A and type D orbits are shown, with black and red dots marking equal time and proper time intervals $\Delta t = \Delta \tau = 5M $. As expected, the dots are more widely spaced when closer to the black hole, and the lengths of sections corresponding to proper time intervals $\Delta \tau$ are longer than those corresponding to time intervals $\Delta t$ (which is also clear in the $t=t(r)$ and $\tau=\tau(r)$ plots of \reff{fig:outA} and \ref{fig:outD}, where $t(r)>\tau(r)$ for all $r$). Note that in case of type D orbit, the dots are plotted only for one orbital period, while the orbit is plotted for 3 periods to show the periapsis precession.

In order to compare the efficiency and accuracy of the analytical expression \refe{resiteva} for $t$ to a direct numerical integration, equation \refe{eq:times1} has been integrated using fourth-order Runge-Kutta method with adaptive step-size control \cite{numericalC}. The elliptic integrals in equation \refe{resiteva} were calculated by Carlson's algorithm \cite{Carlson}, while the Jacobi elliptic functions in \refe{eq:lambda_chi} were from \cite{numericalC}.

For type A orbit ($E=1.01$, $\lt=2.2$) the integration limits were $r_{min}= 6.15313M \approx\rp$ and $r_{max}=50M$, while for type D orbits ($E=0.9704$, $\lt= 1.888$), the limits were $r_{min}=5.04581M\approx \rp$ and $r_{max}=25.436 M\sim\ra$. In both cases, the numerical integration fails, if $r$ gets too close to either $\rp$ or $\ra$ since these are the zeroes of the polynomial $P(u)$ in \refe{eq:times1}. Taking e.g. $r_{min}=\rp(1+10^{-8})$ and $r_{max}=\ra(1-10^{-8})$ and thus avoiding the divergence,\footnote{Consequently, the periapsis and apoapsis passage times have to be calculated in a different manner.} it turns out that numerical integration is $\sim50-80$ times slower than analytical solution \refe{resiteva}. In addition, the relative error $\delta t/t$  is $\sim 4$ orders of magnitude and $\sim 2$ orders of magnitude larger for numerical integration than for analytical solution \refe{resiteva} in case of type A and type D orbits, respectively.

It should be also noted that some additional effort is required when numerically integrating \refe{eq:times1}: if the orbit passes either $\ra$ or $\rp$, e.g. a type D orbit spans many periods (or even just one!), or a type A orbit passes the periapsis, some book-keeping of periapsis and apoapsis passages has to be done in order to obtain the correct solution, e.g. by adding the correct number of half-periods. If using analytical solution, no such additional work is necessary, since \refe{resiteva} is essentially expressed with an angle along the orbit.

%% file: typeB.tex
The polynomial $P(u)$ can be factorized as $P(u) = (u - u_1)(u^2 + pu +q)$, where $u_1 < 0$ is the only real root (see \reff{fig:tipi}). The coefficients $p$, $q$, and the root $u_1$ are expressed as \cite{nickalls,andreja}:
\begin{subequations}
\label{sub:par_b1}
\begin{align}
	\ddd & = (\alpha - \sqrt{D})^{1/3}\label{ddb}\\
	 \ddt & = (\alpha + \sqrt{D})^{1/3}\\
  	u_1 & = \frac{1}{3}\Bigl( 1+ \ddd + \ddt \Bigr)\label{u1mB}\\
	p & = u_1 - 1\label{eq:pb}\\
    q & = -b + p u_1\label{qb}\ .
\end{align}
\end{subequations}

Using the substitution \cite{elipticni}
\beq
	u(\chi)  = u_1 + \sqrt{u_1^2 + p u_1 + q} \tan^2 \frac{\chi}{2} \label{eq:u_chi_b}\ ,
\eq
equations \refee{eq:orbit_diff}{eq:times} are transformed into Legendre form of elliptic integrals and integrated to obtain orbital variables $\lambda$, $t$, and $\tau$ as functions of $\chi$:
\begin{align}
	\lambda (\chi) & = n \left( \elF (\chi|m) - \elF(\chi_{\infty}|m)\right)\label{eq:lambda_chi_b}\\
\begin{split}
	t(\chi) & = 2a\left\{
						\frac{1}{k_1^2}\left[ 
											\alpha_2\left(	n_1^2	\left(	1 + \frac{1}{k_1}	\right)^2 
											\left( 2 + \frac{n_1^2 - m}{n_1(m-n_1)}\right) - 1\right) -\alpha_1 n_1 \left(1 + \frac{1}{k_1} \right)	
										\right]\Pi (n_1;\chi|m) \right. \\
				 		&\quad + \frac{(n_1 -1)(1+k_1)}{2\sqrt{|n_1 -m|}}\left[ 
																			\alpha_2\left(	n_1	\left(	1 + \frac{1}{k_1}	\right)
																			\left( 1 + \frac{1 - m}{n_1-m}\right) - 2\right) -\alpha_1	
																		\right] \ln |x_1|\\
						&\quad + \left[
										2\alpha_2 (n_1 -1)\left(1 + \frac{1}{k_1} \right) - \frac{\alpha_1}{k_1} - \frac{\alpha_3}{k_2}
								 \right]\elF(\chi|m)\\
						&\quad + \frac{\alpha_2}{m-n_1}\left(\frac{n_1}{k_1}\left(1+\frac{1}{k_1}\right)\right)^2\left[
										\elE(\chi|m) -\frac{n_1}{k_1}\frac{\sin\chi \sqrt{1 - m\sin^2 \chi}}{1 -n_1\sin^2\chi}(1+k_1\cos\chi)\right]\\
						&\quad + \alpha_3 (1 -n_2)\left(1 +\frac{1}{k_2}\right)
								\left[ \Pi(n_2;\chi|m) + \frac{k_2}{\sqrt{n_2-m}}\ln|x_2|\right]
				   \Biggr\} 
	\label{resitev_b}
\end{split}\\
\begin{split}
	\tau (\chi) & = \frac{2\alpha_2}{\lt k_1^2}\left\{
				\left( 	\left(\frac{k_1}{k_1 - 1}\right)^2 \left(1 - \frac{m(n_1 - 1)}{n_1(n_1-m)}	\right) 
				-1\right)\Pi (n_1;\chi|m) \right . \\
				& + \frac{n_1(1+k_1)}{2\sqrt{|n_1-m|}}\left( \frac{k_1}{k_1 - 1}\frac{n_1+1-2m}{n_1-m}
				 - 2\right)\ln |x_1| + \frac{2k_1}{k_1 - 1}\elF(\chi|m) \\
				& + \left . \frac{1}{m-n_1}	\left(\frac{k_1}{k_1 - 1}\right)^2\left[ \elE(\chi|m)
					-\frac{n_1}{k_1}\frac{\sin\chi \sqrt{1 - m\sin^2 \chi}}{1 -n_1\sin^2\chi}(1+k_1\cos\chi)\right]\right\}\ , \label{eq:tau_b}
\end{split}
\end{align}
where:
\begin{subequations}
\label{sub:par_b2}
\begin{align}
	m & = \frac{1}{2}\left(1 - \frac{u_1 + p/2}{\sqrt{u_1^2 + p u_1 + q}} \right)\label{mb}\\
    n & = (u_1^2 + p u_1 + q)^{-1/4}\label{eq:n_b}\\
	\alpha_1 & = \frac{n^3}{n^2 u_1+1} \label{alfa1}
	\esp
	\alpha_2 = \frac{n^5}{(n^2 u_1+1)^2}
	\esp
	\alpha_3 = \frac{n^3}{n^2(1-u_1)-1}\\
	k_1 & = \frac{1 - u_1 n^2}{1 + u_1 n^2}
	\esp
	k_2 = \frac{1+(1-u_1)n^2}{1-(1-u_1)n^2}\\
	n_1 & = \frac{k_1^2}{k_1^2-1}
	\esp
	n_2 = \frac{k_2^2}{k_2^2-1}\\
	x_1 & = \frac{
		          \sqrt{n_1-m}\sin\chi+\sqrt{1-m\sin^2 \chi}
		          }{
			  \sqrt{n_1-m}\sin\chi-\sqrt{1-m\sin^2 \chi}
		          }\label{eq:x1_b}\\
	x_2 & = \frac{
		          \sqrt{n_2-m}\sin\chi+\sqrt{1-m\sin^2 \chi}
		          }{
			  \sqrt{n_2-m}\sin\chi-\sqrt{1-m\sin^2 \chi}
		          } \label{x2} \ .
\end{align}
\end{subequations}
Inverting \refe{eq:lambda_chi_b} by $\chi (\lambda) = \am (	\elF (\chi_{\infty}|m) + \lambda/n\ | m )$ and using \refe{eq:u_chi_b} it is straightforward to obtain the following form of the orbit equation:
\beq
u(\lambda) = u_1 + \frac{1}{n^2}\hspace{1pt}
        \frac{ 
	      1-\cn\bigl(\elF(\chi_{\infty}|m) + \frac{\lambda}{n}|m
	      \bigr) 
	     }{
	      1+\cn\bigl(\elF(\chi_{\infty}|m) + \frac{\lambda}{n}|m
	      \bigr) 
	      } . \label{orbitaBm}
\eq

\begin{figure}
\includegraphics[width=\textwidth]{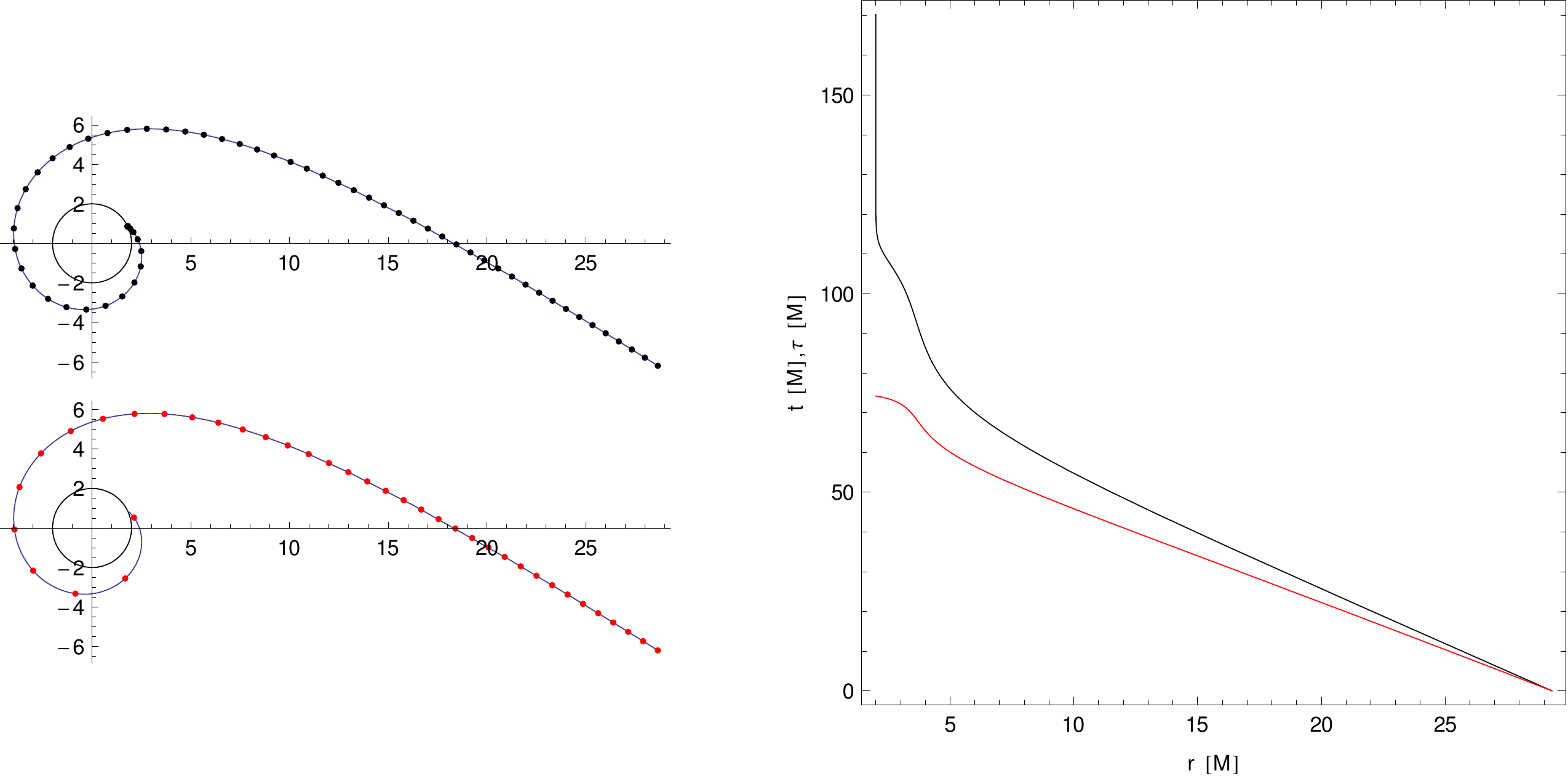}
\caption{Left: Orbit of type B with $E=1.06$ and $\lt=2.2$. Black and red dots correspond to points at time intervals $\Delta t=2 M$ and proper time intervals $\Delta\tau = 2M$, respectively. The black circle represents the Schwarzschild radius. Right: Time (black) and proper time (red) as a function of coordinate $r$, measured from the initial point at $r=29 M$.}
\label{fig:outB}
\end{figure}

The values of $\chi$ are in the interval $\chi \in (\chi_{BH},\chi_{\infty})$, where  $\chi_{BH}=\arccos\frac{1-n^2(1-u_1)}{1+n^2(1-u_1)}$ and $\chi_{\infty}=\arccos\frac{1+n^2u_1}{1-n^2u_1}$. Since neither periapsis nor apoapsis exist for this type of orbits, the value of $\lambda$ is measured from the direction toward infinity, i.e. $\lambda = 0$ at $r\rightarrow \infty$ and the values of $\lambda$ are in the interval $\lambda/n \in (\elF(\chi_{BH}|m) - \elF(\chi_{\infty}|m),0)$. Additionally, it is clear from equations \refe{eq:times1} and \refe{eq:times} that while time $t$ diverges as $r\rightarrow 2M$, proper time $\tau$ remains finite (see \reff{fig:outB}).

In \reff{fig:outB}, an example of the solution for a type B orbit is shown, with black and red dots marking equal time and proper time intervals $\Delta t = \Delta \tau = 2M $. As in previous case, the dots are more widely spaced when closer to the black hole, and the lengths of sections corresponding to proper time intervals $\Delta \tau$ are longer than those corresponding to time intervals $\Delta t$. However, since $t\rightarrow\infty$ when $r\rightarrow 2M$, the black dots start to concentrate at $r\sim 2M$, while the red ones remain distinctly separated. This is also visible in the $t=t(r)$ and $\tau=\tau(r)$ plots of \reff{fig:outB}, where $t(\sim2M)$ diverges and $\tau(\sim 2M)$ has a finite value.

The efficiency and accuracy of the analytical expression \refe{resitev_b} for $t$ compared to a direct numerical integration of \refe{eq:times1} has been done using the same methods as in the previous case. For type B orbit ($E=1.06$, $\lt=2.2$) the integration limits were $r_{min}= 2.0001M$ and $r_{max}=100M$. The numerical integration is $\sim20$ times slower than analytical solution \refe{resitev_b} and the relative error $\delta t/t$ is $\sim 2$ orders of magnitude larger for numerical integration than for analytical solution \refe{resitev_b}.

%% file: typeC.tex
Since type \emph{C} orbits exist for both $D>0$ and $D\leq0$, two different sets of parameters are introduced: If $D > 0$, use the parameters \refe{sub:par_b1} and \refe{sub:par_b2} 
for type \emph{B} orbits. If $D \leq 0$, use the parameters from \refee{dd}{sub:par_a} 
to calculate $p = -(u_2 + u_3)$ and $q = u_2 u_3$, and use them in \refee{eq:pb}{qb} and \refe{sub:par_b2}. In both cases, the root $u_1$ can be associated to the radius of apoapsis $r_a=2M/u_1$. Also, if $m>n_1$, do the following substitution in equations \refe{resitev_b} and \refe{eq:tau_b}:
\beq
	\ln|x_1| \rightarrow 2\arctan (y_1)\ ,
\eq
where $y_1$ is
\beq
y_1 = \frac{\sqrt{m-n_1}\sin\chi}{\sqrt{1-m\sin^2\chi}}\ .
\eq
This substitution is necessary because if $m>n_1$ then $x_1$ becomes complex, so it is more convenient to use the relation $\ln((1+ix)/(1-ix))= 2\Arth(ix)= 2i \arctan(x)$, where the imaginary unit $i$ cancels out with $i$ from $\sqrt{m-n_1}$ in front of the $ln$ term.

While the solutions for $u$, $t$, and $\tau$ are the same as for type \emph{B}, the solution for $\lambda$ is
\beq
	\lambda (\chi) = n \elF (\chi|m)\label{eq:lambda_chi_c}\ ,
\eq
i.e. use equations \refee{eq:u_chi_b}{orbitaBm} with the above replacements. Inverting \refe{eq:lambda_chi_c} by $\chi (\lambda) = \am (	\lambda/n\ | m )$ and using \refe{eq:u_chi_b} it is straightforward to obtain the following form of the orbit equation:
\beq
u(\lambda) = u_1 + \frac{1}{n^2}\hspace{1pt}
        \frac{ 
	      1-\cn\bigl(\frac{\lambda}{n}|m
	      \bigr) 
	     }{
	      1+\cn\bigl(\frac{\lambda}{n}|m
	      \bigr) 
	      } . \label{orbitaCm}
\eq

The limits for $\lambda$ are $\lambda/n \in \{ -\elF(\chi_{BH}\vert m),\elF(\chi_{BH}\vert m)\}$, where $\chi_{BH}=\arccos\frac{1-n^2(1-u_1)}{1+n^2(1-u_1)}$. Note that in this case, the values of $\lambda$ and $\chi$ at apoapsis are $\lambda=\chi=0$. In case of type C orbits it is also true that for $r\rightarrow 2M$, time $t$ diverges and proper time $\tau$ remains finite.

\begin{figure}
\includegraphics[width=\textwidth]{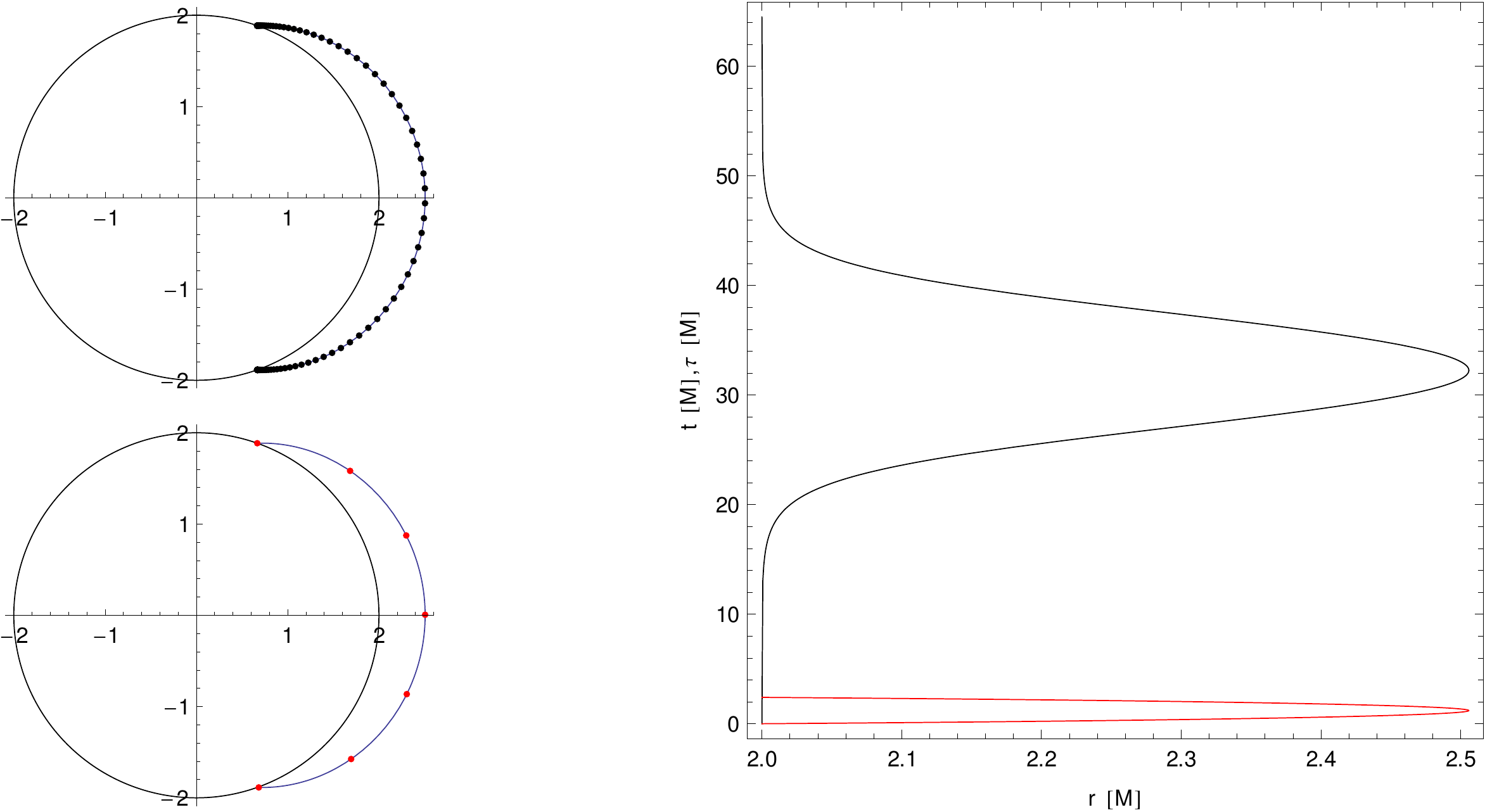}
\caption{Left: Orbit of type B with $E=1.1$ and $\lt=2.8$. Black and red dots correspond to points at time intervals $\Delta t=0.4 M$ and proper time intervals $\Delta\tau = 0.4M$, respectively. The black circle represents the Schwarzschild radius. Right: Time (black) and proper time (red) as a function of coordinate $r$, measured from the initial point at $r=2.0001 M$.}
\label{fig:outC}
\end{figure}

In \reff{fig:outC}, an example of the solution for a type C orbit is shown, with black and red dots marking equal time and proper time intervals $\Delta t = \Delta \tau = 0.4M $. As in case of type B orbit, since $t\rightarrow\infty$ when $r\rightarrow 2M$, the black dots start to concentrate at $r\sim 2M$, while the red ones remain distinctly separated. This is also visible in the $t=t(r)$ and $\tau=\tau(r)$ plots of \reff{fig:outB}, where $t(\sim2M)$ diverges and $\tau(\sim 2M)$ has a finite value. Note that since the orbit is always very close to the black hole, the difference between $t$ and $\tau$ is huge, so the number of $\Delta \tau$ intervals is much smaller than the number of $\Delta t$ intervals.

The efficiency and accuracy of the analytical expression \refe{resitev_b} for type C orbits compared to a direct numerical integration of \refe{eq:times1} has been done using the same methods as in previous cases. For type C orbit ($E=1.1$, $\lt=2.8$) the integration limits were $r_{min}= 2.0001M$ and $r_{max}=2.50581839M\approx\ra$. As in the case of type A and D orbits, the numerical integration fails, if $r$ gets too close to $\ra$. Taking e.g. $r_{max}=\ra(1-10^{-8})$ to avoid the divergence, it turns out that numerical integration is $\sim270$ times slower than analytical solution \refe{resitev_b} and the relative error $\delta t/t$  is $\sim 2$ orders of magnitude larger for numerical integration than for analytical solution \refe{resitev_b} for type C orbit. Also, similarly as in case of type A and D orbits, if the orbit passes the apoapsis $\ra$, this has to be done taken into account only if doing numerical integration of \refe{eq:times1}.

%% file: conclusion.tex
In this paper, the analytical solutions of the orbit equation for time-like geodesics in Schwarzschild space-time are presented in a very straightforward way. The orbits are classified into four types according to the roots of polynomial $P(u)$. This classification is also presented in a more intuitive way, i.e. according to the effective potential and orbital energy. The four orbit types are: type \emph{A} - scattering orbits with both endpoints at infinity, type \emph{B} - plunging orbits with one end at infinity and the other behind the horizon, type \emph{C} -  near orbits with both ends behind the horizon of the black hole, and type \emph{D} - bound orbits. The analytical solutions are expressed with Jacobi elliptic functions where the true anomaly is the only parameter.

The analytical solutions for time and proper time for all four orbit types are also presented here and are expressed as functions of one parameter $\chi$. A simple relation between $\chi$ and true anomaly $\lambda$ is given for all four types.

Since these analytical solutions for time and proper time are expressed with elliptic integrals, which can be numerically calculated very efficiently and accurately either with Landen transformations \cite{elipticni} or Carlson's algorithms \cite{Carlson}, they can be very useful in particular for modelling dynamical phenomena near black holes. These solutions have been in fact already successfully used together with light-like solutions \cite{2005PhRvD..72j4024C} in modelling tidal disruption of low-mass satellites around black holes \cite{2009A&A...496..307K} and quasi-periodic oscillations from X-ray binaries \cite{2009AIPC.1126..367G}. Although the motivation for this work comes from black hole physics, the method was selected due to its performance \cite{2010gfps.confE..19D} also for investigation of a relativistic approach to Galileo Global Navigation Satellite System \cite{2011AdSpR..47..370D}.